\documentstyle[12pt]{article}
\begin{document}
\vskip .7in
\begin{center}
\Large{A Model for the Three Lepton Decay Mode of the Proton}
\vskip .7in
\large{Biswajoy Brahmachari\footnote{Address after
October 1994 : International Center for Theoretical Physics,
I-34100 Trieste, Italy}\\
Theory Group,\\
Physical Research Laboratory,\\
Ahmedabad - 380009, India.
\\
\vskip .2in
Patrick J. O'Donnell \\
Physics Department,\\
University of Toronto,\\
Toronto, Ontario M5S 1A7, Canada.\\
\vskip .1in
and \\
\vskip .1in
Utpal Sarkar\footnote{{\it On sabbatical leave from} Theory Group,
Physical Research Laboratory, Ahmedabad 380 009, India }\\
Institut f\"{u}r Physik,\\
Universit\"{a}t Dortmund,\\
D-44221 Dortmund, Germany
}
\vskip .4in
\baselineskip 16pt
\begin{abstract}
An  extension  of  the  left--right   symmetric  model  has  been
constructed  which gives in a natural way the three  lepton decay
modes of the proton which have been  suggested as an  explanation
for  the  atmospheric   neutrino   anomaly.  We  write  down  the
potential which after minimization gives the proper choice of the
Higgs  spectrum.  With this  Higgs  spectrum  we then  study  the
evolution of the gauge coupling  constants and point out that for
consistency one has to include effects of gravity.
\end{abstract}
\end{center}
\newpage
\baselineskip 18pt
\section{Introduction}

One expects to see produced in the atmosphere  twice as many muon
neutrinos   as  electron   neutrinos   since   detectors   cannot
distinguish   between   neutrinos  and   antineutrinos.  The  two
water--Cerenkov  detectors give a result which is a factor of two
smaller for the ratio  $R=N(\nu_\mu)/N(\nu_e)$,  a ratio in which
many  systematic   uncertainties  are  expected  to  cancel.  The
results  are  $$R_{obs}/R_{MC}=0.60  \pm 0.07 \pm 0.05$$ from the
Kamiokande  experiment  \cite{atm1}  (based on 6.1 Kton year) and
$$R_{obs}/R_{MC}=0.54    \pm   0.05   \pm    0.12$$    from   IMB
\cite{atm2}(based  on 7.7 Kton Year).  The  experiments  look for
``contained''  single  prong events which are caused by neutrinos
with energies  below 2 GeV.  The ratio $R$ is estimated  from the
relative rates of sharply defined single rings  (muon--like)  and
diffused  single rings  (electron--like).  (Other reported values
\cite{SP} for this ratio are:  Frejus \cite{atm3} $0.87 \pm 0.21$
(1.56 Kton  Year),  NUSEX  \cite{atm4}  $0.99 \pm 0.40$ (0.4 Kton
year), SOUDAN II \cite{atm5} $0.69 \pm 0.19$ (1 Kton year)).

Although  this  atmospheric   neutrino   anomaly  has  a  popular
explanation   within   the   neutrino    oscillation    framework
\cite{atmrev}, there is an alternative explanation based on three
lepton decays of the proton  \cite{mann}.  The single ring events
have been analysed within the proton decay  interpretation  where
it is argued that, if the proton  decays into a positron  and two
neutrinos  with a lifetime  of $\tau(P  \rightarrow  e^+ \nu \nu)
\sim 4 \times 10^{31}$ years, then the excess  observed  electron
events  could be due to  proton  decay  events  \cite{mann}.  The
lifetime for this particular decay mode of the proton \cite{taup}
is  consistent  with the  present  limit  \cite{tauprot}  for the
expected  dominant decay mode of the proton  $\tau(P  \rightarrow
e^+ \pi^\circ) > 5 \times 10^{32}$ yr.  The possibility that this
particular decay mode might dominate over other usual decay modes
was  considered  earlier on  general  grounds  \cite{pss,pati,os}
where  it was  pointed  out that it is  difficult  to have  light
neutrinos in the final decay product.

In most models the left--handed neutrinos are light and the right
handed--neutrinos are heavy.  Thus the decay modes are restricted
to
\begin{equation}
P \to e^+ \nu_L \nu_L \;\;\;\; {\rm or} \;\;\;\; P \to e^+
{\nu_L}^c  {\nu_L} \;\;\;\; {\rm or} \;\;\;\; P \to e^+
{\nu_L}^c  {\nu_L}^c.  \label{dkmod}
\end{equation}
Recently \cite{os} it has been pointed out that these decay modes
are allowed in the  framework  of certain  left--right  symmetric
models.  In this paper we  construct  an explicit  model in which
three  lepton decay mode of the proton is the  dominant  one.  We
write down the general  form of the  potential  and show that the
minima of the potential are  consistent  with the choice of Higgs
structure.  We then  study the  evolution  of the gauge  coupling
constants  including  non-renormalizable  interactions  which may
arise from the Planck scale physics \cite{hidim,utpal}.

\newpage
\section{A Proton Decay Mechanism}

We work  in the  framework  of the  left--right  symmetric  model
\cite{pss,pati,lrm} and start with the symmetry breaking chain
\begin{eqnarray}
&SU(4) \times SU(2)_L \times
SU(2)_R \left[ \equiv G_{PS} \right]& \nonumber \\
&& \nonumber \\
{M_{PS} \atop \longrightarrow} & SU(3)_c \times SU(2)_L
\times SU(2)_R \times U(1)_{(B-L)}  \left[ \equiv G_{LR}
\right]& \nonumber  \\
&& \nonumber\\
{M_R \atop \longrightarrow} & SU(3)_c \times SU(2)_L  \times
U(1)_Y \left[ \equiv G_{std} \right] &\nonumber \\
&& \nonumber\\
{M_W \atop \longrightarrow} & SU(3)_c \times U(1)_{em}.& \nonumber
\end{eqnarray}

In the minimal  left--right  symmetric model  \cite{pss,pati,lrm}
the Higgs  scalars  consist of the  following  fields.  The group
$G_{PS}$ is broken by the vacuum expectation value ($vev$) of the
field $H$ which transforms as ({\bf 15},{\bf 1},{\bf 1}) under
the  group  $G_{PS}$.  The right  handed  group is broken  by the
$vev$ of a right handed  triplet Higgs field  $\Delta_R  \equiv $
({\bf 1},{\bf  1},{\bf 3},-2)  $\subset$  ({\bf 10},{\bf  1},{\bf
3}).  By left--right  parity this will imply the existence of the
left  handed  triplet  field  $\Delta_L  \equiv  $ ({\bf  1},{\bf
3},{\bf 1}, -2) $\subset$ ({\bf 10},{\bf 3},{\bf 1}), which gives
Majorana  mass to the  left--handed  neutrinos  and  whose  $vev$
should be $\le 1$ GeV.  (Where there are four  numbers, the first
three correspond to the representations of $SU(3)_c\times SU(2)_L
\times SU(2)_R$ while the last shows the $U(1)$ quantum numbers).

Finally the electroweak symmetry breaking takes place through the
$vev$ of a doublet  scalar  field  $\phi  \equiv$  ({\bf  1},{\bf
2},{\bf 2},0)  $\subset$  ({\bf 1},{\bf  2},{\bf 2}).  This field
$\phi$ also gives masses to the  fermions.  However this does not
reproduce  the right  quark--lepton  mass  ratios.  For the right
magnitude of the quark--lepton mass ratios we require yet another
field $\xi \equiv$ ({\bf 1},{\bf  2},{\bf 2},0)  $\subset$  ({\bf
15},{\bf 2},{\bf 2}) \cite{pss}.  The $SU(3)_c$ singlet component
of  this  field  $\xi$,  which  acquires   $vev$,  has  different
Clebsch--Gordon  coefficients  for the  $SU(3)_c$  and the $U(1)$
part of $SU(4)$.  Hence they  contribute  to the quark and lepton
masses  with  different   coefficients.  As  a  result,  suitable
combinations   of  $\phi$  and  $\xi$  can  reproduce  the  right
quark--lepton mass ratios.

It was pointed out in ref \cite{os} that with this minimal scalar
content it is possible to get the decay mode  required to explain
the atmospheric neutrino problem.  For this we need the $SU(3)_c$
color  triplet  components  of the fields  $\Delta_L$  and $\xi$,
which we  represent  by  $\Delta_L^3$  and $\xi^3$  respectively.
($\Delta_L$ is ({\bf 10},{\bf  3},{\bf 1}) under $G_{PS}$ and the
({\bf 1},{\bf 3},{\bf 1},-2) component acquires a $vev$; the {\bf
10} representation of $SU(4)$  decomposes under $SU(3)_c$ as {\bf
6 + 3 +1}.  Similarly,  the {\bf 15}  representation  of  $SU(4)$
decomposes under $SU(3)_c$ as ${\bf 8 + 3 + \bar{3} + 1}$).  Then
the Yukawa couplings,

\begin{equation}
{\cal L}_{Yuk} = f_{ql} (\overline{{q_{L}}^c} l_{L})
\Delta_{L}^{3^*} + f_{dl} (\overline{{d_{R}}^c} {l_{L}}^c) \xi^{3^*} \,
\end{equation}
and the quartic scalar coupling,
\begin{equation} {\cal L}_{s} =
\lambda^{pr} \Delta_{L}^{3} \Delta_{L}^{3} \xi^{3} \xi^{1} \label{quartic}
\end{equation}
give the $(B - L)$ conserving proton decay $P \rightarrow {e_L}^+
\nu_L  {\nu_L}^c$  through the diagram of figure 1.  This diagram
will  also  give,  with  equal  probability,  the decay  mode, $P
\rightarrow {\mu_L}^+ \nu_L {\nu_L}^c$.

Such a proton  decay  mechanism  will give equal  number of sharp
single  rings  (muon--like   events)  and  diffuse  single  rings
(electron--like   events).  Since   the   proton   decay   events
contribute to both electron-- and muon--like events this seems to
imply that the  reduction of the ratio $R$ cannot be explained by
proton  decay  events.  However, the weighted  average of the two
processes   with   ratios    $R(proton\;    decay)   =   1$   and
$R(atmospheric\;  neutrino) = 2$ (the theoretical  expected ratio
for the muon--to--electron  events if they have their origin only
from  the   atmospheric   neutrinos)  can  in  fact  explain  the
atmospheric neutrino anomaly.

To see  this  we note  the  observed  numbers  of  electron--like
[muon--like] events $n_e(obs)$  [$n_\mu(obs)$] are the sum of the
numbers of electrons [muons] produced by the atmospheric electron
[muon] neutrinos $\nu_e$  [$\nu_\mu$]  through  scattering inside
the detector $n_e(atm)$ [$n_\mu(atm)$] and from the decays of the
protons  into $e^+ \nu_L  {\nu_L}^c$  [$\mu^+  \nu_L  {\nu_L}^c$]
inside  the  detectors  $n_e(prot)$   [$n_\mu(prot)$].  That  is,
$$R_{obs}  =  {n_\mu(obs)   \over   n_e(obs)}  =   {n_\mu(atm)  +
n_\mu(prot)  \over n_e(atm) + n_e(prot)} \sim 0.6 R_{MC} \sim 1.2
.$$

In ref  \cite{mann}  it was assumed  that the proton  decays into
$e^+ \nu_L {\nu_L}^c$ (and not muons), {\it i.e.}, $n_\mu(prot) =
0$.  They found that, by doing a Monte Carlo simulation to obtain
the proton  lifetime, the atmospheric  neutrino  anomaly could be
achieved with a proton lifetime of $\tau_p \sim 4 \times 10^{31}$
years.  In the above relation this corresponds to  $n_e(prot)\sim
(2/3) n_e(atm)\sim (1/3) n_\mu(atm)$.

In the present scenario the proton decays into both electrons and
muons  so  that   $n_e(prot)\sim   n_\mu(prot)$.  Thus   for  the
explanation  of  the  atmospheric  neutrino  anomaly  we  require
$n_e(prot)\sim 4 n_e(atm)\sim 2 n_\mu(atm)$.  Since the number of
proton decays is increased to give the same $R_{obs}$  there will
be a reduction in the proton  lifetime by a factor of 6.  Thus in
this  scenario we can explain the  atmospheric  neutrino  anomaly
with a proton lifetime  $\tau_p \sim (2/3) \times 10^{31}$ years,
which is still  consistent  with  present  experiments  on proton
decay.

The amplitude for the process is given by
\begin{equation}
{\cal A} = \displaystyle  \frac{ \lambda^{pr}  f_{ql}^2 f_{dl} \langle \xi^1
\rangle }{ m_{\xi^3}^2  m_{\Delta^3}^4} .
\end{equation}
where,  $\langle \xi^1 \rangle = \langle \phi \rangle = 250$ GeV,
$\lambda^{pr}$ is the strength of the quartic coupling defined in
Eq.  (\ref{quartic})  and the  $f_{ql}$,  $f_{dl}$ are the Yukawa
coupling constants.

Then, taking reasonable  values for the quartic and the quadratic
Yukawa  coupling  parameters,  $\lambda^{pr}  \sim  10^{-2}$  and
$f\sim  10^{-3}$, say, requires  $m_{\xi^3}$ and $ m_{\Delta^3} $
to be relatively light.  For the proton decay mode $P \rightarrow
{e_L}^+  \nu_L  {\nu_L}^c$  to be $10^{31}$  years to explain the
atmospheric neutrino anomaly, it has been argued  \cite{pss,pati}
that the mass  $m_{\xi^3}$  can be as light as about a TeV, which
requires $m_{\Delta^3} \sim $ few TeV.  This can also be achieved
naturally \cite{os}.

\newpage
\section{A General Left--Right Symmetric Potential}

We now concentrate on the masses $m_{\xi^3}$ and  $m_{\Delta^3}$.
In earlier references  \cite{pss,pati,os} two new mechanisms were
proposed  which could give rise to  appropriate  masses for these
fields.  Here we check  the  consistency  of these two  different
mechanisms when the complete potential with all the scalar fields
is written and  minimized.  First we describe the two  mechanisms
which keep these two fields $m_{\xi^3}$ and $m_{\Delta^3}$ light.

For light $m_{\xi^3}$ it was argued \cite{pss,pati} that if there
exists a field  $\xi^\prime  \equiv ({\bf  15},{\bf  2},{\bf 2})$
which can mix with the field  $\xi$,  then fine tuning can give a
large  mass  to  one   combination   of  the  fields   $\xi$  and
$\xi^\prime$  and keep the other  combination  with a light mass.
However,  this will also keep the  masses of the color  octet and
the color singlet light, which is  undesirable  from the point of
view of evolution of the gauge coupling  constants.  This problem
is avoided  \cite{pati} if instead of $\xi^\prime$ we introduce a
field $\chi \equiv ({\bf 6},{\bf 2},{\bf 2})$ under $G_{PS}$.  If
the  symmetry  group  $G_{PS}$ is  embedded in the unified  group
$SO(10)$  then this field is  contained  in a {\bf  54}--plet  of
$SO(10)$,  which is required to break the  symmetry  of the large
group.  The mixing of the field $\chi$ with $\xi$ can then give a
mass  matrix  which may be fine tuned to give only a light  color
triplet field.  We shall discuss the details of this mechanism at
a later stage.  In the rest of the article we shall use the field
$\chi$ and not $\xi^\prime$.

For the field $\Delta^3$ to remain light we have to alter the way
in which the left--right  symmetric model gets broken,  although,
as we shall  see  later,  this  particular  method  will make the
mechanism in the last paragraph  consistent with the minimization
of the  general  potential.  For  this  purpose  we  introduce  a
singlet field $\eta \equiv$ ({\bf 1},{\bf 1},{\bf 1},0) $\subset$
({\bf  1},{\bf  1},{\bf  1}), to  break  the  left--right  parity
(usually this is referred to as $D$--parity) at a different scale
from the left--right  symmetry  breaking scale $M_R$  \cite{par}.
This field transforms under $D$ as $\eta \to - \eta$.  The scalar
and  the  fermionic   fields   transform  under   $D$--parity  as
$\Delta_{L,R} \to \Delta_{R,L}$  and $\psi_{L,R} \to \psi_{R,L}$,
while $\phi$ and $\xi$ stay the same.  Then with the field $\eta$
we can add new terms to the lagrangian,

\begin{equation}
{\cal L}_{\eta \Delta} = -M_{\eta} \eta (\Delta_R^\dagger \Delta_R
  - \Delta_L^\dagger \Delta_L) - \lambda_{\eta}  \eta^2
  (\Delta_L^\dagger  \Delta_L + \Delta_R^\dagger \Delta_R) \label{nd}
\end{equation}

In theories where the triplet Higgs breaks $D$--parity along with
$SU(2)_R$ we have  $m_{\Delta_L} =  m_{\Delta_R}$.  The masses of
the fields  $\Delta_L$  and  $\Delta_R$  are not be the same when
$D$--parity  is  broken  by the $vev$ of the field  $\eta$.  When
$\eta$ gets a non--zero vacuum expectation  value, the masses are
given by,
$$  m_{\Delta_L}^2  =  m_\Delta^2 -M_{\eta}  \langle \eta  \rangle
+  \lambda_{\eta}  {\langle  \eta
\rangle }^2 \;\; {\rm and} \;\;  m_{\Delta_R}^2  =  m_\Delta^2  +
M_{\eta}  \langle \eta  \rangle +  \lambda_{\eta}  {\langle  \eta
\rangle  }^2  $$
where  $m_\Delta$ is the mass at which the  left--right  symmetry
breaking  of  $SU(2)_R$  is  broken  spontaneously.  So,  in  the
absence of the $\eta$ field, both these $\Delta$ fields will have
mass $\sim  m_\Delta$.  With $<\eta>$  present the  parameters in
the three terms can be tuned to make $m_{\Delta_L}$  vanish.  The
field  $\Delta_L$  will then  acquire  mass of the order of a TeV
from radiative corrections.  The same sets of parameter will also
make $m_{\Delta_R}$ heavy and lead to a solution
\begin{equation}
\langle \eta \rangle  \sim \langle  \Delta_R  \rangle \gg \langle
\Delta_L   \rangle\;  {\rm  and}   \;   M_\eta   \approx
m_{\Delta_R}  \approx \langle \Delta_R \rangle \approx m_{\Delta}
\; {\rm and} \; m_{\Delta_L} \ll \langle \Delta_R \rangle
\label{mass}
\end{equation}
Thus we can  have  $m_{\Delta^3} \sim m_{\Delta_L} \sim $ few TeV
even when  $m_{\Delta_R} \sim \langle  \Delta_R  \rangle$ is as
large as  $10^{10}$ GeV.

We now  write  the most  general  potential  with all the  fields
present  in the  minimal  left--right  symmetric  model  with the
additional  fields $\xi, \chi $ and the $D$--parity  odd--singlet
field $\eta$.  We then show the two  mechanisms  required to keep
the color triplet fields light are consistent  with the minima of
the potential.  To simplify the expression we define,
\begin{eqnarray}
\phi_1 & \equiv  &\phi~~;~~
\phi_2  \equiv  \tau_2 \phi^{*}_1 \tau_2~~;~~\xi_1  \equiv  \xi~~;~~
\xi_2  \equiv  \tau_2 \xi^{*}_1 \tau_2 \nonumber
\end{eqnarray}
\noindent The most general potential with all the fields is,
\begin{eqnarray}
V(\phi_1,\phi_2,{\Delta_L},{\Delta_R},\xi_1,\xi_2,{\eta},\chi) &=&
V_\phi+V_\Delta+V_\eta+V_\xi+V_{\chi}\nonumber \\
&&+V_{\eta \phi}+V_{\eta \Delta}+
V_{\Delta \phi}+V_{\phi\xi}+V_{\Delta\xi}+V_{\eta\xi}+V_{\chi\xi}\nonumber \\
\end{eqnarray}
where the different terms in this expression are given by,
\begin{eqnarray}
&&\nonumber\\
V_\phi&=& -\sum_{i,j}\mu^2_{ij}~tr(\phi ^{\dagger}_i
\phi_j)+\sum_{i,j,k,l}\lambda_{ijkl}~tr(\phi ^{\dagger}_i
\phi_j)~tr( \phi ^{\dagger}_k \phi_l)\nonumber\\
&&+\sum_{i,j,k,l}\lambda^\prime_{ijkl}~tr(\phi ^{\dagger}_i \phi_j \phi
^{\dagger}_k \phi_l)\nonumber\\
{}~&&~\nonumber\\
V_\Delta &=&-\mu^2~[tr(\Delta ^{\dagger}_L \Delta_L)+tr(\Delta^{\dagger}_R
\Delta_R)] +
\rho_1~[tr(\Delta ^{\dagger}_L \Delta_L)^2+tr(\Delta ^{\dagger}_R
\Delta_R)^2]\nonumber\\
&& \nonumber \\
&&+\rho_2~[tr (\Delta^{\dagger}_L \Delta_L \Delta^{\dagger}_L \Delta_L)
+tr (\Delta^{\dagger}_R \Delta_R \Delta^{\dagger}_R \Delta_R)]
+\rho_3~tr(\Delta ^{\dagger}_L \Delta_L \Delta ^{\dagger}_R
\Delta_R)\nonumber\\
{}~&&~\nonumber\\
V_\eta&=&-\mu^2_\eta~ \eta^2+\beta_\eta~ \eta^4 \nonumber\\
{}~&&~\nonumber\\
V_\xi&=&\sum_{i,j}m^2_{ij}~tr(\xi ^{\dagger}_i
\xi_j)+\sum_{i,j,k,l}~n_{ijkl}~tr(\xi ^{\dagger}_i\xi_j\xi
^{\dagger}_k\xi_l)
+\sum_{i,j,k,l}p_{ijkl}~tr(\xi ^{\dagger}_i\xi_j)~
tr(\xi ^{\dagger}_k\xi_l) \nonumber\\
{}~&&~\nonumber\\
V_\chi&=&M_\chi^2 tr(\chi^\dagger~ \chi)+\lambda^\chi_1
[tr(\chi^\dagger~\chi)]^2 + \lambda^\chi_1 tr(\chi^\dagger~\chi
\chi^\dagger~\chi) \nonumber\\
&&\nonumber\\
V_{\Delta\phi}&=&+\sum_{i,j}\alpha_{ij}~[tr(\Delta ^{\dagger}_L \Delta_L)
+tr(\Delta^{\dagger}_R \Delta_R)]~{tr(\phi ^{\dagger}_i \phi_j) }
+\sum_{i,j} \beta_{ij}~[~tr(\Delta ^{\dagger}_L\Delta_L\phi_i
\phi ^{\dagger}_j) \nonumber\\
& &+tr(\Delta ^{\dagger}_R\Delta_R\phi ^{\dagger}_i
\phi_j)]+\sum_{i,j} \gamma_{ij}~tr(
\Delta^{\dagger}_L\phi_i \Delta_R\phi^\dagger_j) \nonumber\\
{}~&&~\nonumber\\
V_{\eta\Delta}&=& - M_\eta~\eta~[tr(\Delta^{\dagger}_L \Delta_L)-tr(
\Delta ^{\dagger}_R \Delta_R)]+\lambda_\eta~ \eta^2~[tr(\Delta
^{\dagger}_L \Delta_L)+tr(\Delta^{\dagger}_R \Delta_R)] \nonumber\\
{}~&&~\nonumber\\
V_{\eta\phi}&=&\sum_{i,j}\delta_{ij}~\eta^2~tr(\phi ^{\dagger}_i
\phi_j)\nonumber\\
&&\nonumber\\
V_{\phi\xi}&=&\sum_{i,j,k,l}u_{ijkl}~tr(\phi ^{\dagger}_i \phi_j
\xi ^{\dagger}_k \xi_l)+\sum_{i,j,k,l}v_{ijkl}~tr(\phi
^{\dagger}_i \phi_j)~tr(\xi ^{\dagger}_k \xi_l) \nonumber \\
&&\nonumber\\
V_{\Delta\xi}&=&+\sum_{i,j}a_{ij}~[~tr(\Delta ^{\dagger}_L \Delta_L)+tr(\Delta
^{\dagger}_R \Delta_R)]~{tr(\xi ^{\dagger}_i \xi_j) } \nonumber\\
&& +\sum_{i,j} b_{ij}~[~tr(\Delta ^{\dagger}_L\Delta_L\xi_i
\xi ^{\dagger}_j)+tr(\Delta ^{\dagger}_R\Delta_R\xi ^{\dagger}_i
\xi_j)]\nonumber\\
&&+\sum_{i,j} c_{ij}~tr(\Delta^{\dagger}_L\xi_i \Delta_R\xi^\dagger_j)
\nonumber\\
&&+ \sum_{i,j} \lambda^{pr}_{ij} [tr(\Delta_L \Delta_L \xi_i \xi_j)
+ tr(\Delta_R \Delta_R \xi_i \xi_j) + tr(\Delta_L \Delta_R \xi_i \xi_j)]
\nonumber\\
&&\nonumber\\
V_{\eta\xi}&=&\sum_{i,j}d_{ij}~\eta^2~tr(\xi
^{\dagger}_i \xi_j)\nonumber \\
&&\nonumber\\
V_{\chi\xi}&=&P~\eta[tr(\xi\chi\Delta_R)-tr(\xi\chi\Delta_L)]+
M~[tr(\xi\chi\Delta_R)+tr(\xi\chi\Delta_L)]
\nonumber \\ \nonumber
\end{eqnarray}

We  have  not  written  the  $SU(4)$   indices   explicitly   for
simplicity.  For  example, if we include the  $SU(4)$  index, the
term $\rho_2 tr (\Delta^{\dagger}_L  \Delta_L  \Delta^{\dagger}_L
\Delta_L)$  in  our  notation   will  actually  mean  two  terms,
$\rho_2^a   tr(\Delta^{\dagger    \alpha}_L   \Delta_{L   \alpha}
\Delta^{\dagger   \beta}_L   \Delta_{L   \beta})$  and  $\rho_2^b
tr(\Delta^{\dagger  \alpha}_L  \Delta_{L  \beta}  \Delta^{\dagger
\beta}_L   \Delta_{L   \alpha})$.  However,   as   far   as   the
minimization  and the  consistency  of the model is concerned, we
only  have  to  replace  $\rho_2$  by  $(\rho_2^a  +  \rho_2^b)$.
Otherwise  the rest of the  analysis  will be  unaltered.  A more
detailed   analysis  with  explicit   $SU(4)$  indices  will  not
constrain or relax any of the constraints in this model.

The vacuum expectation values ($vev$) of the fields have the
following form:
\begin{eqnarray}
<\phi>&=& \pmatrix{k&0\cr 0& k^{\prime}}~~;~~
<\Delta_L>= \pmatrix{0&0\cr v_L&0}~~; \nonumber \\
<\tilde{\phi}>&=& \pmatrix{k^{\prime} &0\cr 0&k}~~;~~
<\Delta_R>= \pmatrix{0&0\cr v_R&0}~~; \nonumber\\
<\xi>&=&\pmatrix{{\tilde k}&0\cr 0&
{\tilde k}^\prime}~~;~~<\tilde{\xi}>= \pmatrix{{\tilde k}^{\prime}
&0\cr 0&{\tilde k}}~~; \nonumber\\
<\eta> &=& \eta_0~~;~~~~~~<\chi> = 0 \nonumber
\end{eqnarray}

The notation needs some clarification.  For the fields $\phi$ and
$\xi$ we have used the representation in which rows correspond to
the $SU(2)_L$ quantum numbers  $(+\frac{1}{2},~-\frac{1}{2})$ and
columns    correspond   to   the   $SU(2)_R$    quantum   numbers
$(-\frac{1}{2},~+\frac{1}{2})$.  The  field  $\phi$  is a singlet
under  the  group  $SU(4)$  and  hence  it has a  $SU(4)$  matrix
representation  $diag(1,1,1,1)$.  On the  other  hand  the  field
$\xi$  transforms  under  $SU(4)$  as a {\bf 15}  representation.
Under the $SU(3)$  subgroup of $SU(4)$ the {\bf 15} decomposes as
${\bf 8 + 3 + \bar{3} + 1}$.  The $SU(4)$  matrix  representation
of the  singlet is a traceless  diagonal  matrix  which is a unit
matrix  in  the   $SU(3)$   space.  Hence  the   $SU(4)$   matrix
representation  of the  component  of  $\xi$,  which is a singlet
under both the $SU(3)_C$ and the $U(1)$  subgroups of $SU(4)$ and
which   acquires  a  $vev$,  is   $diag(1,1,1,-3)$.  The  $SU(2)$
representations are as above.  Thus these fields $\phi$ and $\xi$
contribute  differently  to the quarks and  leptons  masses,  and
hence a proper  combination  of the two fields  give the  correct
mass  relations  between the quarks and leptons  \cite{pss}.  For
the fields  $\Delta_L$  and  $\Delta_R$  we used the $2 \times 2$
triplet  representations of $SU(2)$.  Thus the components $\tau^1
- i~\tau^2$, which has the isospin $+1$ and hence charge neutral,
acquire  $vev$.  (The  electric  charge  is  $T_{3L}  +  T_{3R} +
(B-L)/2$.  For these  representations  $B-L = -2$.  So the charge
neutral  component should have $T_{3L}$ or $T_{3R} = +1$, meaning
they should contract with $\tau^1 - i~\tau^2 $).

\newpage
\section{Minimization  of the Potential}

It is almost impossible to minimize the potential with respect to
all the  fields  and then  find  the  absolute  minima.  For this
purpose one needs to  simplify  the  problem  considerably.  As a
first  approximation one can extremize the potential with respect
to all  the  fields  and  then  substituting  the  $vev$s  of the
different fields to check if there is any  inconsistency.  In the
minimal left--right  symmetric potential, {\it i.e.,} without the
field $\eta$, there are no linear terms in any field so the usual
practice  is to replace the various  fields by their  $vev$s then
extremize it with respect to these  $vev$s and finally  check for
consistency.  We shall  also  follow the same  procedure,  but we
need to take  care  of the  extra  linear  terms  present  in the
potential.  For these linear terms, we shall afterwards  minimize
the  potential  with respect to those fields which do not acquire
any $vev$.  The vanishing of these derivatives after substituting
for the $vev$s will then impose new  constraints  which also have
to be satisfied.

After the spontaneous  symmetry breaking, when the fields acquire
a $vev$,  the  potential  contains  terms  with $k$,  $k^\prime$,
${\tilde k}$, ${\tilde k}^\prime$, $v_L$ and $v_R$.  We need only
terms involving $v_L$ and $v_R$.  These are given by,
\begin{eqnarray}
V &=&-\mu^2~(v^2_L+v^2_R)+ {\rho \over
4}~(v^4_L+v^4_R)+{\rho^\prime \over 2}~(v^2_Lv^2_R)
+2v_Lv_R[(\gamma_{11} \nonumber\\
&& \nonumber\\
&&+\gamma_{22})kk^\prime+\gamma_{12}(k^2+{k^\prime}^2)]
+(v^2_L+v^2_R)~[(\alpha_{11}+\alpha_{22}+\beta_{11})~k^2\nonumber\\
&& \nonumber\\
&&+(\alpha_{11}+\alpha_{22}+\beta_{22})~{k^\prime}^2
+(4\alpha_{12}+2\beta_{12})~kk^\prime]\nonumber\\
&& \nonumber\\
&&-M_\eta~\eta_0~(v^2_L-v^2_R)+ \lambda_\eta~ \eta^2_0~(v^2_L+v^2_R)
\nonumber\\
&& \nonumber\\
&&+(v^2_L+v^2_R)~[(a_{11}+a_{22}+b_{11}+\lambda^{pr}_{11})~\tilde{k}^2
+(a_{11}+a_{22}+b_{22}+\lambda^{pr}_{22})~{\tilde{k^\prime}}^2\nonumber\\
&& \nonumber\\
&&+(4a_{12}+b_{12}+\lambda^{pr}_{12})~\tilde{k}\tilde{k^\prime}]
+ 2v_Lv_R[(c_{11}
+c_{22}+\lambda^{pr}_{11}+\lambda^{pr}_{22})\tilde{k}
\tilde{k^\prime} \nonumber\\
&& \nonumber\\
&&+(c_{12}+\lambda^{pr}_{12})(\tilde{k}^2+\tilde{k^\prime}^2)]
\end{eqnarray}
where we have defined the new parameters
$\rho=4(\rho_1+\rho_2)$ and  $\rho^\prime=2 \rho_3$.

The  minimization  of this potential  gives a constraint on $v_L$
and $v_R$.  Instead of minimizing  this potential with respect to
the fields $v_L$ and $v_R$ separately, we consider a combination,
$  {\partial  V \over  \partial  v_L}  v_R -  {\partial  V  \over
\partial  v_R} v_L = 0$, which gives a relation  among the fields
$v_L$ and $v_R$
\begin{equation}
v_Lv_R={\beta_1 k^2+ \beta_2 \tilde{k}^2 \over
[\rho-\rho^\prime -{4 M_\eta \eta_0 \over (v^2_L-v^2_R)}]}. \label{ss}
\end{equation}
where,  $\beta_1 = 2  \gamma_{12};$  and  $\beta_2  = 2 (c_{12} +
\lambda^{pr}_{12})$  and we  assumed  $k'<<k $ and  $\tilde{k}'<<
\tilde{k}$.  This  allows us to have a very  tiny  $vev$  for the
left--handed  triplet field $\Delta_L$ while keeping the $vev$ of
the  right--handed  triplet field $\Delta_R$ very large.  This is
in agreement with what we required in Eq.  (\ref{mass}).

Now  consider  the  linear  terms   involving   $\chi$  given  by
$V_{\chi\xi}$.  These terms allow for the correct mixing  between
the color triplet components of the fields $\xi$ and $\chi$.  The
mass matrix for $\xi^3$ and $\chi^3$ is now given by
\begin{equation}
{\cal M} = \pmatrix{a & b \cr c & d}
\end{equation}
where, $a = m^2 $, $d = M_\chi^2$ and $b = c = (P \eta_0 + M) v_R
$, and where we assumed  all  $m_{ij}$s  are equal to $m$.  If we
now fine tune parameters to make ${\rm det}{\cal M} = 0$, $i.e.$,
$$ (m M_\chi)^2 = (P \eta_0 + M)^2  v_R^2$$  then one of the mass
eigenvalues is zero.  This fine tuning  requires that $M$ must be
negative  and $(P  \eta_0 + M)$ be very  small and  negative.  In
fact,  $|P  \eta_0 + M|$ has to be of the  order of $\sim  {M_W^2
\over M_R}$.  This massless field will get a mass of the order of
$\sim$ TeV after  radiative  corrections  during the  electroweak
symmetry breaking are included.

The terms  linear in  $\chi$  required  for this  mechanism  have
another  effect  which  was  not  transparent  when  we  did  the
minimization  of the potential  with respect to the $vev$s of the
various  fields.  If we first  minimize with respect to the field
$\chi$ and then  substitute  for the $vev$s of the various fields
(which  is  not   usually   done  since  that   complicates   the
calculation), then there is an additional constraint,
\begin{equation}
v_L={P    \eta_0+M   \over   P   \eta_0-M}v_R
\end{equation}

This means that to satisfy Eq.  (\ref{ss}), we require $|P \eta_0
+ M| << |P \eta_0 - M| $.  In other  words, for $v_L \sim  {M_W^2
\over  M_R}$,  we need  $\left|  {P  \eta_0+M  \over P  \eta_0-M}
\right| \sim {M_W^2 \over  M_R^2}$.  This is consistent  with the
fine tuning  used to keep the color  triplet  component  of $\xi$
light, for which $|P \eta_0 + M| \sim {M_W^2  \over M_R}$ and $|P
\eta_0 - M| \sim M_R$.  This would not have been  possible if the
field  $\eta$  were not  present.  For  example, in the  original
paper  \cite{pati}  where the field $\chi$ was introduced and the
field $\eta$ was not  required,  this method would have led to an
inconsistency.  In the absence of the field  $\eta$  minimization
of the  potential  with  respect to the field  $\chi$  would have
given a constraint $v_L = - v_R$, which is inconsistent  with the
LEP data.  Here we need the field $\eta$ to keep the left--handed
color triplet light and the fine tuning makes it consistent.

We now turn to the  question of light  $\Delta^3$.  As  mentioned
earlier, to have $M_\Delta \sim$ few TeV, we require the coupling
of $\eta$ and  $\Delta$ as in Eq.  (\ref{nd}).  Also, in the most
general potential the only term which contributes at the level of
$\eta_0 \sim M_{\Delta_R}$ is $V_{\eta \Delta}$, which is exactly
the same as in (Eq.  \ref{nd}).  Thus the  field  $\Delta^3  \sim
\Delta_L$  can  be  massless  at  that  level.  Then  during  the
electroweak phase transition this will again acquire mass through
radiative  corrections  of the  order of a few TeV.  This is more
natural   in   supersymmetric   theories   where  the   radiative
corrections  induce  mass of the order of  supersymmetry  breaking
scale, which are usually of the order of a few TeV.

The mechanism just mentioned to make $\Delta^3$ light has one
drawback. It makes the other components of $\Delta_L$ also very
light. For example, the $\Delta^6$ can now mediate $n-\bar{n}$
oscillation, which has to be suppressed. This problem is similar
to the doublet-triplet splitting of any other grand unified models.
In the present scenario we assume that although this field is light
the coupling of $\Delta^6$ is very small, which can make the model
safe. However, this is not the best choice and if one can find some
good solution to the doublet-triplet splitting in other GUTs, then
one has to incorporate
the same mechanism here in future. With our present assumption that
the Yukawa coupling of $\Delta^6$ is very small, we now have to check
the consequences of these fields in the evolution of the gauge coupling
constants.

Finally,  we point  out  that  the  quartic  coupling  (given  by
$V_{\Delta \xi}$) required for generating the required diagram is
also present in the general potential.

\newpage
\section{Evolution of the Coupling Constants}

There have to be many light  scalars for the present  scenario to
work.  These scalars may destabilize the unification of the gauge
coupling constant at the unification  scale.  In the evolution of
the gauge coupling constant with these Higgs scalars included and
with  the  mass  scales  as  above  it  is  impossible   to  have
unification  of  the  gauge  coupling  constants  using  the  LEP
constraints   \cite{lep}  on   $\sin^\theta_w$   and  $\alpha_s$.
However higher dimensional  operators, which might originate from
Planck scale physics such as quantum gravity or  compactification
of  Kaluza-Klein  theories or Superstring  theories, can save the
situation\cite{hidim,utpal}.  Thus if the three lepton decay mode
of the proton survives all of the experimental tests, then we may
have  an  indication   that  Planck  scale  physics  is  actually
modifying the boundary conditions of the gauge coupling constants
near the unification scale.

In our  analysis we include the effect of the  non-renormalizable
terms arising from Planck scale physics from the beginning  using
the notation and method of reference \cite{utpal}.  We write down
the  generalized  renormalization  group  equations  in which the
Planck  scale  effects  are  parametrized  in terms of four extra
parameters.  We recover the usual relations  between the coupling
constants  in  the  absence  of  gravity  by  setting  the  extra
parameters   to  zero.  It  is  when  we  do  this  we  obtain  a
contradiction and the equations fail to provide unification.

The evolution of gauge  coupling  constants with the energy scale
is governed by  Renormalization  Group Equations  (RGE).  Here we
consider the RGE in one loop approximation i.e.  the gauge fields
fermionic  fields  and  the  scalar  fields   contribute  to  the
evolution  of the  gauge  couplings  via  one  loop  graphs  only
\cite{gut}.  In  this  approximation  the  renormalization  group
equation takes the form,
$$
\mu \frac{{\rm d} \alpha_i (\mu)}{{\rm d} \mu} = 2 b_i \alpha_i^2 (\mu)
$$
where, $\alpha_i = \frac{g_i^2 }{4 \pi}$ and
the beta function is given in the following generic form.
\begin{equation}
b = { 1\over 4 \pi} \left[ -{11 \over 3} N+ {4 \over 3}
n_f + { 1\over 6} T_s  \right] \label{beta}
\end{equation}
Here $N = 1, 2, 3 \, { \rm or } \, 4$, the number of neutrinos is
always 3 and the  scalars  take on the  values  discussed  below.
Since there are a large  number of scalar  fields  present in our
model  their   contribution  will  be  substantial   despite  the
suppression by a factor of $6$ in the beta function of the scalar
term ($T_s$).  We list the scalar fields that  contribute  to the
RGE at  different  energy  scales  in  Table.  \ref{table1}.  For
simplification  of notation  we write $M_c$ for  $M_{PS}$ in this
section.

We embed the  Pati-Salam  group  $G_{PS}$ into a larger GUT group
SO(10).  The  $SO(10)$  symmetry is broken by a {\bf  54}-plet of
Higgs  field  $\Sigma$  at the scale  $M_U$.  The  $\Sigma$  is a
traceless  symmetric  field of the $SO(10)$ and the {\it vev}s of
$\Sigma$ which mediates this symmetry breaking are given by,
\begin{equation}
        \langle \Sigma \rangle = \displaystyle
\frac{1}{\sqrt{30}}\:\: \Sigma_0 \:\: {\rm diag}(1, 1, 1, 1, 1,
1, -\frac{3}{2}, -\frac{3}{2}, -\frac{3}{2}, -\frac{3}{2}).
\end{equation}
where,
$\Sigma_0=\sqrt{6 \over 5\pi\alpha_G} M_U$ and
$\alpha_G=g_0^2/4\pi$ is the GUT coupling.
The $vev$ of a 45-plet field $H$ breaks the symmetry group
$G_{PS}$ to $G_{LR}$,
\begin{equation}
\langle H \rangle = \displaystyle \frac{1}{\sqrt{12}\:
i} \:\: H_0 \: \pmatrix{ 0_{33} & 1_{33} & 0_{34} \cr -1_{33} & 0_{33} &
0_{34} \cr 0_{43} & 0_{43} & 0_{44} }
\end{equation}
where, $0_{mn}$ is a $m \times n$ null matrix and $1_{mm}$ is a
$m \times m$ unit matrix.

The {\bf (1,1,1)}  component of the {\bf 54}-plet  field $\Sigma$
breaks the  $SO(10)$  group at the scale $M_U$ and hence does not
affect  the  RGE.  The  {\bf  (15,1,1)}  component  of  the  {\bf
45}-plet field $H$ breaks  $G_{PS}$ and so contributes to the RGE
between the scale $M_c$ and $M_U$.  The color singlet part of the
{\bf (10,1,3)}  component of the {\bf 126}-plet field  $\Delta_R$
breaks the $G_{LR}$ symmetry.
Then by extended survival hypothesis the
color singlet component contribute to the RGE between the scale
$M_R$ to $M_c$ and all the components {\bf (10,1,3)} contribute
between the scales $M_c$ and $M_U$ \cite{gut}.
On the  other  hand our  proposed  mechanism,  allows  all of the
components  of the field  $\Delta_L  \equiv$  {\bf  (10,3,1)}  to
remain light and  contribute  to the RGE at all energies  between
$M_W$ and $M_U$.  The bidoublet  {\bf  (1,2,2)}  field breaks the
electroweak symmetry group and contributes at all energies to the
RGE .  For the correct quark-lepton mass relation we also require
the  bidoublet  color  singlet  component  of {\bf  (15,2,2)}  to
acquire a {\it vev} at the electroweak  scale.  For the potential
we have, the {\bf (6,2,2)}  field will mix with the color triplet
and  anti-triplet  components of the {\bf (15,2,2)} field and one
combination of these color triplet and  anti-triplet  fields will
remain  light.  The  color  singlet  and one  combination  of the
triplet and  anti-triplet  component will then  contribute to the
RGE at all  energies,  while the other  combination  of the color
triplet and  anti-triplet  will become heavy and contribute  only
between the energies $M_c$ and $M_U$.

\begin{table}[htb]
\begin{center}
\begin{tabular}{|c||c||c|}
\hline
$M_U \rightarrow M_{c}$&$M_{c} \rightarrow M_{R}$&$M_{R}
\rightarrow M_W$\\
\hline
(1,1,1)&&\\
\hline
(15,1,1)&&\\
\hline
({10},1,3)&(1,1,3,$\sqrt{3 \over 2}$)&\\
\hline
(10,3,1)&(6,3,1,$-\sqrt{3 \over 2}$)+(3,3,1,$\sqrt{1 \over 6}$)
+(1,3,1,$\sqrt{3 \over 2}$) &(6,3,$-\sqrt{3 \over 5}$)
+(3,3, ${ 1\over 3} \sqrt{ 3\over5}$)+(1,3,$\sqrt{3 \over 5}$)\\
\hline
(1,2,2)&(1,2,2,0)&(1,2,${1 \over 2} \sqrt{3 \over 5}$)+(1,2, ${
-1\over 2} \sqrt{ 3\over
5}$) \\
\hline
(15,2,2)&(3,2,2,$\sqrt{2 \over 3}$)+($\bar{3}$,2,2,-$\sqrt{2 \over
3}$)+(1,2,2,0)&(3,2,${1\over 6}\sqrt{ 3\over 5}$)+(3,2,${
7\over 6} \sqrt{ 3\over 5}$)+(1,2, ${ 1\over 2} \sqrt{ 3\over
5}$) \\
&&($\bar{3}$,2, ${ -1\over 6} \sqrt{ 3\over 5}$)+($\bar{3}$,2, ${-7
\over6} \sqrt{ 3\over 5}$)+(1,2, ${ -1\over 2} \sqrt{ 3\over
5}$)\\
\hline
(6,2,2)&& \\
\hline
\end{tabular}
\end{center}
\caption{Higgs  scalars at various symmetry breaking scales.  The
$U(1)$  quantum  numbers are normalized  from their  embedding in
$SO(10)$.}
\label{table1}
\end{table}
The normalization of the U(1) quantum numbers at the right handed
breaking scale is fixed by the relation,
\[
Y = \sqrt{3 \over 5} T^3_R + \sqrt{2 \over 5} Y_{B-L}.
\]
Combining  results of Eq.  \ref{beta} and Table.  \ref{table1} it
is easy to write  down the  following  explicit  form of the beta
functions that regulate the  evolutionary  behaviour of the gauge
couplings at various energy scales.  We assume that the number of
fermion generations is three.
\begin{table}[htb]
\begin{center}
\begin{tabular}{|c||c||c|}
\hline
$M_U \rightarrow M_{c}$&$M_{c} \rightarrow M_{R}$&$M_{R}
\rightarrow M_z$\\
\hline
\hline
&&\\
$b_{4uc}$= -11/3 &
$b_{3cr}$= -29/6 &
$b_{3crw}$ = -29/6 \\
&&\\
$b_{2Luc}$= 11/3 &
$b_{2Lcr}$= 4/3 &
$b_{2Lrw}$ = 4/3 \\
&&\\
$b_{2Ruc}$= 11/3 &
$b_{2Rcr}$= 3/2 &
$b_{1Yrw}$= 49/6 \\
&&\\
&$b_{1B-Lcr}$= 13 &\\
&&\\
\hline
\hline
\end{tabular}
\end{center}
\caption{The  modified beta functions  ($\widetilde{b_N}  = 4 \pi
b_N$) for the various groups at different  energy scales.  In the
table we use the notation $b_Nxy$, where $N$ represents the group
($N$ for $SU(N)$ and $1S$ for  $U(1)_S$)  and $xy$ means the beta
functions within the scales $M_x$ and $M_y$.}
\label{table2}
\end{table}

We consider both symmetry breaking scales, $M_U$ and $M_c$, to be
very large so that Planck scale  effects are not  negligible.  We
start with the renormalizable $SO(10)$ invariant lagrangian,
\begin{equation}
L=-{1 \over 2} {\rm Tr} (F_{\mu\nu} F^{\mu\nu})
\end{equation}
and then include the non-renormalizable  higher dimensional terms
which have their  origin in Planck  scale  physics.  We  consider
only terms of dimension 5 and 6, given by
\begin{equation}
L^{(5)}=-{1 \over 2} {\eta^{(1)} \over M_{Pl}} {\rm Tr} (F_{\mu\nu}
\phi F^{\mu\nu})
\end{equation}
\begin{eqnarray}
L^{(6)}=&-{1 \over 2} {1 \over M_{Pl}^2} \biggl\lbrack
\eta_a^{(2)}\lbrace {\rm Tr} (F_{\mu\nu} \phi^2 F^{\mu\nu}) +
{\rm Tr} (F_{\mu\nu} \phi F^{\mu\nu}\phi)\rbrace + \nonumber \\
&\eta_b^{(2)} {\rm Tr}(\phi^2){\rm Tr} (F_{\mu\nu} F^{\mu\nu}) +
\eta_c^{(2)} {\rm Tr} (F^{\mu\nu}\phi) {\rm Tr} (F_{\mu\nu}
\phi) \biggr\rbrack
\end{eqnarray}
where  $\eta^{(n)}$  are  dimensional  couplings  of  the  higher
dimensional  operators.  When any Higgs  scalar  $\phi$  acquires
$vev$  $\phi_0$, these  operators  induce  effective  dimension 4
terms modifying the boundary conditions at the scale $\phi_0$.

The symmetry  breaking at $M_U$ shifts the boundary  condition of
the  $SU(4)$  coupling  constant  with  respect  to  the  $SU(2)$
couplings  whereas the {\it vev}s of the {\bf 45}-plet  field $H$
contribute  to the  relative  couplings  of the  $SU(3)$  and the
$U(1)$ constants.  The $G_{PS}$ invariant  effective  lagrangian,
modified by these higher dimensional operators, is given by,
\[
-\frac{1}{2}(1+\epsilon_4) \: {\rm Tr} (F_{\mu
\nu}^{(4)}\:F^{(4)\mu \nu}) -\frac{1}{2}(1+\epsilon_{2}) \: {\rm
Tr} (F_{\mu \nu}^{(2L)}\:F^{(2L)\mu \nu})
\]
\begin{equation}
-\frac{1}{2}(1+\epsilon_{2}) \: {\rm Tr} (F_{\mu
\nu}^{(2R)}\:F^{(2R)\mu \nu}  )
\end{equation}
where, $$\epsilon_4 = \epsilon^{(1)} + \epsilon^{(2)}_a +
\frac{1}{2} \epsilon^{(2)}_b $$
$$\epsilon_2 = - \frac{3}{2} \epsilon^{(1)} + \frac{9}{4}
\epsilon^{(2)}_a + \frac{1}{2} \epsilon^{(2)}_b. $$ and
$$
\epsilon_i^{(n)}= \biggl\lbrack \biggl\lbrace{1 \over
25\pi\alpha_G} \biggr\rbrace^{1 \over 2} {M_U  \over
M_{Pl}}\biggr\rbrack^n \eta_i^{(n)} . $$
Then the usual  $G_{PS}$  lagrangian  can be  recovered  with the
modified coupling constants,
\begin{eqnarray}
g_4^2(M_U)&=& \bar{g}_4^2(M_U) (1 + \epsilon_4)^{-1}\nonumber\\
g_{2L}^2(M_U)&=& \bar{g}_{2L}^2(M_U) (1 + \epsilon_2)^{-1}\nonumber\\
g_{2R}^2(M_U)&=& \bar{g}_{2R}^2(M_U) (1 + \epsilon_2)^{-1}
\end{eqnarray}
where the $\bar{g}_i$  are the coupling  constants in the absence
of  the  nonrenormalizable  terms  and  $g_i$  are  the  physical
coupling   constants   that  evolve  below  $M_U$.  The  modified
boundary condition then reads,
\begin{equation}
 {g}_4^2(M_U) (1 + \epsilon_4) =  {g}_{2L}^2(M_U) (1 +
\epsilon_2) =  {g}_{2R}^2(M_U) (1 + \epsilon_2) = g_0^2. \label{eqn24}
\end{equation}

At $M_c$ the  symmetry  group  $SU(4)_c$  breaks down to $SU(3)_c
\times U(1)_{B-L}$ when the {\bf (15,1,1)}  component of the {\bf
45}-plet of Higgs field $H$  acquires a {\it vev}.  The  $SU(3)_c
\times U(1)_{B-L}$ invariant lagrangian is given by,
$$
-\frac{1}{2}(1+\epsilon_3^\prime) \: {\rm Tr} (F_{\mu
\nu}^{(3)}\:F^{(3)\mu \nu}) -\frac{1}{2}(1+\epsilon_{1}^\prime)
\: {\rm Tr} (F_{\mu \nu}^{(1)}\:F^{(1)\mu \nu})
$$
where, $$ \epsilon_3^\prime = \epsilon_a^{\prime(2)} + 12
\epsilon_b^{\prime(2)} $$  $$ \epsilon_1^\prime =
7 \epsilon_a^{\prime(2)} + 12 \epsilon_b^{\prime(2)} + 12
\epsilon_c^{\prime(2)}. $$ and $$
\epsilon_i^{\prime(2)} = \displaystyle \frac{\eta_i^{\prime(2)}
\phi_0^2}{ 24 M_{Pl}^2} = \left[ \frac{1}{20 \pi \alpha_4}
\biggl\lbrack\frac{M_I}{M_{Pl}}\biggr\rbrack^{2} \right] \eta_i^{\prime(2)}
$$
where,  $i = a, b, c$.  Then  the  boundary  condition  at  $M_c$
becomes,
$$
g_{1(B-L)}^2(M_c) (1 + \epsilon_1^\prime) = g_{3c}^2(M_c) (1 +
\epsilon_3^\prime) = g_4^2(M_c).
$$
The matching  conditions  at the scale $M_R$ are not  modified by
the Planck scale effects and are given by,
\begin{eqnarray}
g_{1Y}^{-2}(M_R) & = & {3 \over 5} g_{2R}^{-2} (M_R) +
{2 \over 5} g_{1 (B-L)}^{-2} (M_R) \nonumber \\
g_{2L}^{-2}(M_R) & = &  g_{2R}^{-2} (M_R)  \label{eq2}
\end{eqnarray}

Using   the   above   boundary   conditions   and  the  one  loop
renormalization   group   equation   the   unification   coupling
$\alpha_U$  can be related to the three  couplings  at the W mass
scale ($M_W$) through the following  relations  \cite{utpal}  (we
have defined $m_{ij}={\rm ln} {M_i \over M_j}$),

\begin{eqnarray}
\alpha^{-1}_y(M_W) &=&\alpha^{-1}_G~( 1 + 3/5 \epsilon_2 + 2/5 \epsilon_4 +
2/5 (\epsilon^\prime(1+\epsilon_4))) + (6/5~b_{2ruc}+ 4/5
(1+\epsilon^\prime_1)~b_{4uc})~m_{uc} \nonumber \\
&&+ (6/5~b_{2rcr}+
4/5~b_{1blcr})~m_{cr} + 2 b_{1yrw}~m_{rw} \nonumber\\
\alpha^{-1}_2(M_W) &=& \alpha^{-1}_G~(1+\epsilon_2)+ 2~b_{2luc}~m_{uc} + 2~
b_{2lcr}~m_{cr} +2~b_{2lrw}~m_{rw} \nonumber\\
\alpha^{-1}_3(M_W) &=& \alpha^{-1}_U~(1+\epsilon_4) + 2~b_{4uc}~m_{uc} +
2~b_{3cr} ~m_{cr} + 2~b_{3crw}~m_{rw} \nonumber
\end{eqnarray}
We define, $$
A= \alpha^{-1}_Y(M_W)-\alpha^{-1}(M_Z) $$ and $$
B= \alpha^{-1}_{2L} (M_W) + {5 \over 3} \alpha^{-1}_Y(M_Z) - { 8 \over
3} \alpha^{-1}_{3c}(M_W) $$ and relate them to the experimental numbers
through the following equations,
\begin{eqnarray}
\sin^2_{\theta_W} &=& { 3 \over 8} - { 5 \over 8} \alpha A \nonumber\\
1- { 3 \over 8}{ \alpha \over \alpha_s} &=& \alpha B .
\end{eqnarray}

If we now take all the $\epsilon$ s to be zero, then we have
\begin{eqnarray}
m_{rw} &=& -36.7 + 5.7 m_{uc}\nonumber\\
m_{cr} &=&  147.3 -11.2 m_{uc} \ . \nonumber
\end{eqnarray}
There is no solution with positive  $m_{rw}$ and $m_{cr}$ for any
value of $M_{uc}$ with the constraints, $\sin^2 \theta_W=0.2334$,
$\alpha_s=0.12$,  and with the unification scale below the Planck
scale .  In other words this means that if we do not consider the
effect of gravity,  then in the  presence of so many light  Higgs
scalars  it is not  possible  to have  unification  of the  gauge
coupling  constants.  Thus if the three  lepton decay mode of the
proton is the explanation of the  atmospheric  neutrino  problem,
gravity  effects modify the low energy  predictions  of the grand
unified theories.

We now  consider  the effects of the Planck  scale.  For  several
choices of the  $\epsilon$\ s it is possible to have a mass scale
solution which may explain the atmospheric  neutrino problem.  To
demonstrate this we present a few representative solutions in table
2.  The light  scalars  contribute  from the scale $M_z \sim 100$
GeV in the RG  equation.  The  unification  scale  $M_U$  and the
$G_{PS}$  breaking scale $M_c$ are very close to each other ({\it
i.e.,}  $m_{uc}={\rm  ln} {M_U \over M_c}$ is very  small) and is
considered  here to be  around  $M_U \sim M_c \sim  10^{18}$  GeV
({\it  i.e.,}  $m_{rw} + m_{cr} + m_{uc} = 39$); the right handed
breaking  scale is around  $M_R \sim  10^{13}$  GeV ({\it  i.e.,}
$m_{rw}  \sim 28$).  With  these  values of the mass  scales  the
three lepton decay mode of the proton would be the most  dominant
decay mode (with  $\tau(P  \to e^+ \nu \nu) \sim  10^{31}$  yrs).
Since  the  unification  scale is quite  high  now,  conventional
proton decay modes are very much suppressed (with $\tau(P \to e^+
\pi^\circ)  \sim  10^{39}$  yrs.).  Thus  even  though  the three
lepton decay mode of the proton explains the atmospheric neutrino
anomaly there is no conflict with the  non-observation  of proton
decay in other experiments.
\begin{table}[htb]
\begin{center}
\[
\begin{array}{|c||c||c||c||c||c||c||c|}
\hline
\epsilon^\prime_1&\epsilon_2 &\epsilon^\prime_3 &\epsilon_4
&\alpha^{-1}_G&m_{rw}&m_{cr}& m_{uc} \\ \hline
-1&-.75& 1& -1 & 55& 28.32 & 9.73&.93\\
-1& -.75& 1& -1&  57&   28.04&   10.34&   .61\\
-1& -.75& 1& -1&  59 &  27.75 &  10.95 &  .28\\
-.99& -.75& 1& -1&   59&   27.76&   10.95&   .28\\
-.90& -.75& 1& -1&   59 &   27.76&   10.94&   .28\\
-.80& -.75& 1&  -1&   59 &  27.77 &  10.93 &  .28\\
\hline
\end{array}
\]
\end{center}
\caption{Allowed ranges of parameters for unification}
\end{table}

\newpage
\section{Conclusions}

We have presented an extension of the left--right symmetric model
where  the most  dominant  proton  decay  mode is  through  three
leptons.  The  lifetime  for this decay  mode is large  enough to
explain the atmospheric  neutrino anomaly.  We have minimized the
complete potential to check the consistency of the model.  In the
end we have  carried  out a  renormalization  group  analysis  to
estimate  the mass  scales of the model.  We have shown that when
the gravity  induced  effects coming from the Plank scale physics
are included in the renormalization group analysis, the couplings
unify; an estimation of the several mass scales of the model then
becomes possible.

{\bf Acknowledgement}

We would like to thank Drs.  Debajyoti Choudhuri and N.K.  Mandal
for  many  critical   comments  on  the  manuscript  and  helpful
discussions.  The work of US was  supported by a fellowship  from
the Alexander  von Humboldt  Foundation  and that of PJO'D by the
Natural Sciences and Engineering Council of Canada.

\newpage
\begin{figure}[htb]
\vskip 6.5in\relax\noindent\hskip -.5in\relax{\includegraphics{prfg.ps}}
\caption{Diagram giving $P \rightarrow {e_L}^+ \nu_L {\nu_L}^c$}
\end{figure}
\end{document}